\documentstyle[sprocl]{article}
\input epsf
\def\be{\begin{equation}}
\def\ee{\end{equation}}
\def\bea{\begin{eqnarray}}
\def\eea{\end{eqnarray}}

\def\kpc{\,{\rm kpc}}

\def\cmm2{{\,\rm cm^{-2}}}
\def\cm2{{\,{\rm cm}^2}}
\def\cmm3{{\,{\rm cm}^{-3}}}
\def\gcmm3{{\,{\rm g\,cm^{-3}}}}
\def\kms{\,{\rm km\,s^{-1}}}

\def\la{\mathrel{\mathpalette\fun <}}

\def\fun#1#2{\lower3.6pt\vbox{\baselineskip0pt\lineskip.9pt
  \ialign{$\mathsurround=0pt#1\hfil##\hfil$\crcr#2\crcr\sim\crcr}}}

\begin{document}
\title{MICROLENSING AND THE COMPOSITION OF THE GALACTIC HALO }
\author{ Evalyn Gates$^{\rm a,b}$, Geza Gyuk$^{\rm c}$, and M.S. Turner$^{\rm a,d,e}$ }
\address{$^a$Department of Astronomy \& Astrophysics, Enrico Fermi
Institute, University of Chicago, Chicago IL
60637-1433\\
$^b$Adler Planetarium, Chicago, IL 60605\\
$^c$Scuola Internazionale Superiore di Studi Avanzati, Trieste, Italy\\
$^d$Department of Physics, Enrico Fermi Institute, University of Chicago, Chicago IL 60637-1433\\
$^e$NASA/Fermilab Astrophysics Center, Fermi National Accelerator
Laboratory, Batavia, IL 60510-0500}
\maketitle
\abstracts{By means of extensive galactic modeling we study the
implications of the more than 100 microlensing events that
have now been observed for the composition of the dark halo of
the Galaxy.  Based on the currently published data, including the 2nd year MACHO results,  the halo MACHO fraction is less than 60\% in most models and the likelihood function for the halo MACHO fraction peaks around 20\% - 40\%,  consistent with expectations for cold dark matter models.}

Gravitational microlensing provides a valuable tool for probing
the baryonic contribution to the dark matter in the halo of our Galaxy.
However, even with precise knowledge of the optical depths toward the LMC
and bulge, it would still be difficult to interpret the results
because of the large uncertainties in the structure of the
Galaxy. As it is, small number statistics for the LMC lead to a
range of optical depths further complicating the analysis.
Detailed modeling of the Galaxy is essential to drawing reliable
conclusions.

The values of the parameters that describe the components of the
Galaxy are not well determined;  in order to
understand these uncertainties we explore a very wide range of
models that are consistent with all the data that constrain the
Galaxy.
We consider two basic models for the bulge, a triaxial model
with the long axis oriented at an angle of about $10^{\circ}$ with
respect to the line of sight toward the galactic center, and an axisymmetric
 model.  The bulge mass is not well determined, and we take $M_{\rm
Bulge}=(1 - 4) \times 10^{10} M_\odot$.
For the disk component  we consider a double exponential distribution and
take the sum of a ``fixed,'' thin luminous disk and a dark disk
with varying scale lengths $r_d
= 3.5\pm 1 \kpc$, and thicknesses $h=0.3\kpc$, and $1.5\kpc$.
We also consider a model where the projected mass density varies
as the inverse of galactocentric distance.
We constrain the local projected mass density 
of the dark disk to be $10M_\odot\,\le \Sigma_{\rm VAR}
\le 75 M_\odot \,{\rm pc}^{-2}$.  The dark halo is assumed to be comprised of two components,
baryonic and non-baryonic, whose distributions are independent.
We first assume independent isothermal distributions for the MACHOs
and cold dark matter, with core radii varying between 2 and 20$\kpc$.
Since there are indications from both observations \cite{Sackflat}
and CDM simulations\cite{quinn} that halos are significantly
flattened, we
also consider models with an axis ratio $q = 0.4$ (E6 halo)
for both the baryonic and non-baryonic halos.  While flattening
does affect the local halo density significantly, increasing it
by roughly a factor of $1/q$, it does not
affect the halo MACHO fraction significantly \cite{apjlett}.
Finally, we consider the possibility that the MACHOs are not
actually in the halo, but instead, due to dissipation, are more
centrally concentrated in a spheroidal component.  

We then require that the following
observational constraints be satisfied: circular rotation speed at
the solar circle ($r_0 = 8.0\kpc \pm 1\kpc$) $v_c =220\kms\pm
20\kms$; peak-to-trough variation in $v(r)$ between $4\kpc$ and
$18\kpc$ of less than 14\%;
local escape velocity $v_{\rm ESC} > 450\kms$
and circular rotation velocity at $50\kpc$, 
 $180\kms \leq v_c(50\kpc) \leq 280\kms$.
We also impose constraints from microlensing, both toward the
bulge and toward the LMC.  In calculating the optical depth toward the bulge, we consider
lensing of bulge stars by disk, bulge and halo objects; for the
LMC we consider lensing of LMC stars by halo and disk objects.
We adopt the following constraints based upon microlensing data: 
(a)\cite{ogle,machob}
$\tau_{\rm BULGE} \geq 2.0\times 10^{-6}$ and 
(b)\cite{MACHOprl,eros} $0.4\times10^{-7}\le \tau_{\rm LMC} \le 4\times 10^{-7}$.

\begin{figure}[t]
\begin{center}
\leavevmode
\epsfxsize=3.3in \epsfbox{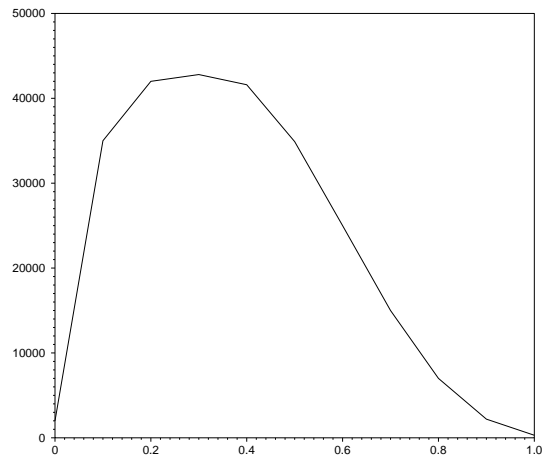}
\end{center}
\caption{The number of viable models as a function of halo
MACHO fraction.  }
\label{fig:halofrac}
\end{figure}

 We summarize here our main results; details of the analysis can be found in ref. 8.

\begin{itemize}

\item The implications of the second year MACHO results for the halo MACHO fraction are shown in Figure~1.   Incorporating the full set of constraints we find that the halo MACHO fraction is less than $60\%$ in most models and peaks at a value of $20\% - 40\%$.   
However, there are                                                                                                                                                                                                                                                                     a small number of allowed models with a halo MACHO fraction greater than $80\%$.   In addition to having a smaller total halo mass, these models all require an optical depth toward the LMC of greater than $2.5 \times 10^{-7}$, and most have  $\tau_{LMC}\ge 3.5\times 10^{-7}$.   

\item Bulge microlensing provides a crucial constraint to galactic
modeling and eliminates many models.  It all but necessitates a
bar of mass at least $2\times 10^{10}M_\odot$ and provides additional evidence
that the bulge is bar-like.  Because of the interplay between
the different components of the Galaxy, the bulge microlensing
optical depth indirectly constrains the MACHO fraction of the halo.
On the other hand, LMC microlensing only constrains the MACHO
fraction of the halo.

\item Viable models with no MACHOs in the halo (where the LMC optical depth is due to a thick disk or spheroidal component population) are difficult unless  $\tau_{LMC}\la 2.0\times 10^{-7}$.
\end{itemize}

\noindent This
work was supported in part by the DOE (at Chicago and Fermilab) and the
NASA (at Fermilab through grant NAG 5-2788).

\end{document}